\documentclass[fleqn,usenatbib,useAMS]{mnras}
\usepackage{graphicx}	
\usepackage{amsmath}	
\usepackage{amssymb}	
\usepackage{multicol}   
\usepackage{bm}		
\usepackage{pdflscape}	
\usepackage{appendix}
\usepackage{ulem}
\usepackage{tcolorbox}
\usepackage{color}
\usepackage{soul}
\usepackage[utf8]{inputenc}
\usepackage{cancel}

\usepackage{multibib}
\newcommand{\beq}{\begin{equation}}
\newcommand{\eeq}{\end{equation}}
\newcommand{\bse}{\begin{subequations}}
\newcommand{\ese}{\end{subequations}}
\newcommand{\bary}{\begin{eqnarray}}
\newcommand{\eary}{\end{eqnarray}}
\newcommand{\bwt}{\begin{widetext}}
\newcommand{\ewt}{\end{widetext}}

\usepackage[T1]{fontenc}
\usepackage{ae,aecompl}

\title{TeV afterglow from GRB 221009A: photohadronic origin?}

\author[S. Sahu et al.]{
Sarira Sahu$^{1}$%
\thanks{Contact e-mail: \href{mailto:sarira@nucleares.unam.mx}{sarira@nucleares.unam.mx}},\ %
B.~Medina-Carrillo$^{2}$%
\thanks{Contact e-mail: \href{mailto:benjamin.medina@cinvestav.mx}{benjamin.medina@cinvestav.mx}},\ %
D. I. Páez-Sánchez$^{3}$%
\thanks{Contact e-mail: \href{mailto:paez@ciencias.unam.mx}{paez@ciencias.unam.mx}},\ %
G.~Sánchez-Colón$^{2}$%
\thanks{Contact e-mail: \href{mailto: gabriel.sanchez@cinvestav.mx}{gabriel.sanchez@cinvestav.mx}},\ %
Subhash Rajpoot$^{4}$%
\thanks{Contact e-mail: \href{mailto: Subhash.Rajpoot@csulb.edu}{Subhash.Rajpoot@csulb.edu}}%
\\
$^{1}$Instituto de Ciencias Nucleares, Universidad Nacional Aut\'onoma de M\'exico,\\ Circuito Exterior S/N, C.U., A.P. 70-543, CDMX 04510, México.\\
$^{2}$Departamento de Física Aplicada, Centro de Investigación y de Estudios Avanzados del IPN, Unidad Mérida.\\ A.P. 73, Cordemex, Mérida, Yucatán 97310, México.\\
$^{3}$Facultad de Ciencias, Universidad Nacional Aut\'onoma de M\'exico, \\ 
Circuito Exterior S/N, C.U., A. P. 70-543, CDMX 04510, México. \\
$^{4}$Department of Physics and Astronomy, California State University,
1250 Bellflower Boulevard, Long Beach, CA 90840, USA.
}

\date{}

\pubyear{2024}

\begin{document}
\label{firstpage}
\pagerange{\pageref{firstpage}--\pageref{lastpage}}
\maketitle

\begin{abstract}
Gamma-ray burst (GRB), GRB 221009A, a long-duration GRB, was observed simultaneously by the Water Cherenkov Detector Array (WCDA) and the Kilometer Squared Array (KM2A) of the Large High Altitude Air Shower Observatory (LHAASO) during the prompt emission and the afterglow periods. Characteristic multi-TeV photons up to 13 TeV were observed in the afterglow phase. The observed very high-energy (VHE) gamma-ray spectra by WCDA and KM2A during different time intervals and in different energy ranges can be explained very well in the context of the photohadronic model with the inclusion of extragalactic background light models. In the photohadronic scenario, interaction of high-energy protons with the synchrotron self-Compton (SSC) photons in the forward shock region of the jet is assumed to be the source of these VHE photons. The observed VHE spectra from the afterglow of GRB 221009A are similar to the VHE gamma-ray spectra observed from the temporary extreme high-energy peaked BL Lac (EHBL), 1ES 2344+514 {\it only} during the 11th and the 12th of August, 2016. Such spectra are new and have been observed for the first time in a GRB.
\end{abstract}

\begin{keywords}
BL Lacertae objects, gamma-ray bursts, relativistic processes
\end{keywords}


\newpage

\section{Introduction}

Gamma-ray bursts are some of the most enigmatic and explosive events in the universe, releasing vast amounts of energy in a fraction of a second. The duration of these bursts of gamma-rays typically last from only a few seconds to a few minutes, and can outshine their entire galaxy in that brief period. The GRBs are classified into two categories, short-duration ($\lesssim 2$ s) and long-duration ($\gtrsim 2$ s)~\citep{1993ApJ...413L.101K}. It is generally believed that the long-duration GRBs are due to the collapse of massive stars and the short-duration GRBs are from the merger of compact objects like neutron stars and/or black holes~\citep{KUMAR20151,BZhang}. Among the thousands of GRBs that have been observed, GRB 221009A, a long-duration GRB,  stands out as a particularly intriguing  event and has captivated astronomers and astrophysicists worldwide due to its unique characteristics~\citep{Williams:2023sfk,Burns:2023oxn}. 

After the initial burst of gamma-rays, there is often an "afterglow" phase, where the burst continues to emit radiation across a broad  frequency range, from optical to GeV energies~\citep{BZhang}. Afterglows from GRBs are studied to gain insight into the nature of the GRBs, their progenitors, different particle acceleration mechanisms and many other properties. Several of the GRBs are also detected at very high-energies (VHE, $> 100$ GeV) in the afterglow era~\citep{MAGIC:2019lau,Abdalla:2019dlr,HESS:2021dbz}. GRBs are detected in keV and MeV energies by space based instruments in the prompt phase~\citep{2004ApJ...611.1005G,Meegan_2009}. Afterwards, the Cherenkov detectors were slew to their positions in the afterglow phases for the VHE observations. Before the observation of GRB 221009A, neither the prompt emission phase nor the transition from the prompt phase to the afterglow phase were observed by the Cherenkov telescopes from the GRBs. On October 9, 2022, GRB 221009A was first observed by the Gamma-Ray Burst Monitor (GBM)~\citep{2022GCN.32636....1V,2022GCN.32642....1L}. The redshift of the GRB was measured to be $z=0.151$~\citep{2022GCN.32648....1D, 2022GCN.32686....1C}. Fortunately, during the triggering of GBM, the position of GRB 221009A was in the field of view of LHAASO. The high sensitivity of LHAASO, accompanied by the very high flux of VHE gamma-rays from GRB 221009A, enabled LHAASO to detect the growth and decay phases of the VHE spectrum. LHAASO-WCDA~\citep{LHAASO:2019qtb} detected more than 64,000 photons in the energy range $\sim$ 200 GeV to $\sim$ 7 TeV within first $\sim 3000$ s after the GBM trigger (trigger time is $T_0$)~\citep{2023Sci...380.1390L}. The temporal profile of the observed VHE spectrum was smooth, with a rapid rise to a peak value, followed by a gradual decay for about 3000 s. To the contrary, the light curves measured by several other gamma-ray detectors, including GBM, during the prompt phase of GRB 221009A are highly variable and multi-pulsed. The contradictory behavior of the MeV and TeV emissions from GRB 221009A suggests that the latter emission does not originate from the prompt phase, but from the forward shock region of the jet. Also, it is believed that LHAASO did not detect TeV emissions during the prompt phase. Probably, the TeV emission in the early stage of the afterglow epoch of GRB 221009A was not contaminated by the prompt phase debris. Hence, the study of early afterglow in the TeV regime is important for understanding the particle acceleration mechanism and emission mechanisms in the GRB jet. However, the possible detection of VHE photons by LHAASO during the prompt phase is also discussed by~\cite{Wang:2023yhx}.

From the temporal profile of the observed VHE spectrum, LHAASO-WCDA has chosen five time intervals to cover the main emission period~\citep{2023Sci...380.1390L}. The observed differential spectra during the five different phases can be explained very well by a power-law with exponential cutoff (PLEC) functions, and the extragalactic background light (EBL) subtracted intrinsic fluxes can be fitted by power-law functions. In the context of the leptonic models, the X-ray afterglow is interpreted as the synchrotron emission from the electrons and the TeV afterglow is interpreted as the SSC emission from the high-energy electrons in the GRB jet~\citep{2023Sci...380.1390L}. Several leptonic, hadronic and leptohadronic models are also used to explain the multiwavelengh emission from GRB 221009A ~\citep{2019ApJ...884..117W,2019ApJ...880L..27D,2023ApJ...943L...2L,2023arXiv231011821W,2023ApJ...947L..14Z}.

Recently, LHAASO-KM2A reported the detection of 142 photons above 3 TeV energies between $T_0+230$ s and $T_0+900$ s~\citep{doi:10.1126/sciadv.adj2778}. For the reconstruction of the VHE events, LHAASO-KM2A uses two different functional fits to the spectra, (i) with a log-parabola (LP) function and (ii) a PLEC. In the LP fit, the maximum energy of the photon is at $17.8^{+7.4}_{-5.1}$ TeV. However, with the PLEC fit, the maximum energy of the photon is $12.2^{+3.5}_{-2.4}$ TeV. From these fits, it is argued that observation of photons $\gtrsim 13$ TeV from a source at $z=0.151$ requires either a relatively transparent Universe for high-energy gamma-rays or the incorporation of new physics, such as Lorentz Invariance Violation~\citep{2022arXiv221007172B,Zhu:2022usw,Li:2022wxc,2023ApJ...942L..21F} or photon-axion oscillation~\citep{2022arXiv221005659G,Troitsky:2022xso,Lin:2022ocj}.

Previously, we have shown that the photohadronic model can explain very well the VHE gamma-ray emission from high-energy peaked BL Lac objects (HBLs)~\citep{2019ApJ...884L..17S,10.1093/mnras/staa023}. It is observed that there are many similarities between the blazar jet and the GRB jet, despite large differences in their masses, bulk Lorentz factors, magnetic fields etc. Thus, in previous studies we have applied the photohadroinc model to explain the VHE gamma-ray spectra of GRBs~\citep{Sahu_2020,Sahu_2022,Sahu:2022gvx}. In this letter, our aim is to use the photohadronic model to interpret, first, the observed VHE spectra of GRB 221009A by LHAASO-WCDA in the five time intervals, and second, the combined VHE spectra of WCDA and KM2A up to $\sim 13$ TeV in the afterglow epoch.

\section{Comparing GRB with Blazar}

The general characteristics of both, blazars and GRBs, and the ranges of their physical parameters, such as bulk Lorentz factor ($\Gamma$), duration of emission, luminosity, baryon loading factor, etc. are discussed in \cite{1999ASPC..161..264D,2005IJMPA..20.6991G,2006smqw.confE..27G,KUMAR20151,BZhang}. From the spectral modeling of the VHE spectra of blazars, a range for the bulk Lorentz factor required is $3 < \Gamma < 50$. On the other hand for GRBs, $\Gamma > 100$ is required. The average luminosity carried by the jet in a blazar is in the range $\sim\,10^{43}-10^{47}\,\mathrm{erg\,s^{-1}}$ and for GRBs, assuming isotropic emission, the total energy can be in the range $10^{50} - 10^{55}$ erg in few seconds.

We have so far observed few GRBs in VHE gamma-rays in the afterglow epochs since 2018. All these VHE spectra are explained by using the SSC process and/or hadronic process. Within the leptonic scenario, the mechanism of VHE gamma-ray production can be explained through the SSC process, where the background seed photons are the synchrotron photons in the jet background or by the inverse Compton (IC) scattering of the relativistic leptons with the photons from the external sources. The TeV afterglow from GRB 221009A is interpreted as the SSC emission from the high-energy electrons in the GRB jet~\citep{2023Sci...380.1390L}. However, it is shown that the  SSC process is inadequate to explain the observed VHE spectrum of GRB 221009A. Thus, the electromagnetic cascading from the interaction of accelerated protons with the external shock environment is proposed as the process to explain this VHE spectrum~\citep{2023arXiv231011821W}.

From these similar features shared by the two, it is not unreasonable to deduce that the particle acceleration mechanisms operating in the blazar jet are equally applicable to the GRB jet~\citep{1995PASP..107..803U, 2011ApJ...740L..21W, 2011ApJ...726L...4W, 2012Sci...338.1445N,2013FrPhy...8..661G, 2016MNRAS.455L...1W}.

It is important to note that the leptonic and the leptohadronic models have large number of parameters. Thus the prediction of these models are limited. On the other hand, the simple photohadronic model with minimal assumptions can explain the VHE spectra from blazars and GRBs very well. So, it is natural to extend the photohadronic model to study the VHE spectra of GRB 221009A.

\section{\label{sec3}The Photohadronic model}

Flaring in VHE gamma-rays is common in blazars and the VHE flare is usually explained using the leptonic models, the hadronic models, and hybrids of both~\citep{2008ApJ...679L...9B,refId0,Giannios:2009pi, Cerruti:2014iwa,2019MNRAS.490.2284M}. 

The photohadronic model is initially used to explain the VHE flaring gamma-ray events from HBLs. In this model, protons are accelerated to very high energies $E_p$ in the inner region of the HBL jet and their differential spectrum is a power-law, $dN_p/dE_p \propto E^{-\alpha}_p$, with the proton spectral index $\alpha \ge 2$~\citep{1993ApJ...416..458D}. The value of $\alpha$ is different for non-relativistic shocks, highly relativistic shocks and oblique relativistic shocks~\citep{2005PhRvL..94k1102K,2012ApJ...745...63S}. In the inner region of the jet, the high-energy protons interact with the background SSC photons, primarily through the $p+\gamma \rightarrow \Delta^+$ process. The necessary condition for the production of $\Delta$-resonance is $E_p \epsilon_{\gamma}=0.32\, \Gamma\, {\cal D}{(1+z)^{-2}} \, {\mathrm{GeV^2}}$, where ${\cal D}$ is the Doppler factor and for blazars ${\cal D}\simeq \Gamma$. Here $\epsilon_{\gamma}$ is the observed background seed photon energy and $z$ is the redshift of the source~\citep{Sahu:2019lwj}. The $\Delta$-resonance decays to charged and neutral pions with $\pi^+\rightarrow e^+{\nu}_e\nu_{\mu}{\bar\nu}_{\mu}$ and $\pi^0\rightarrow\gamma\gamma$ respectively~\citep{PhysRevD.85.043012}. The $\gamma$-rays from $\pi^0$ decay are the observed VHE $\gamma$-rays on Earth. 

The observed VHE gamma-ray flux $F_{\gamma}$ from a source at a redshift $z$ is given by~\citep{2019ApJ...884L..17S}
\begin{equation}
F_{\gamma}(E_{\gamma})=F_{in}(E_\gamma)\, e^{-\tau_{\gamma\gamma}(E_\gamma,z)}
=F_0\, E_{\gamma, TeV} ^{-\delta+3}\, e^{-\tau_{\gamma\gamma}(E_\gamma,z)},
\label{eq:flux}
\end{equation}
where the intrinsic flux (IF) is $F_{in}$ and $E_{\gamma, TeV}$ is the photon energy in TeVs. The normalization factor $F_0$ can be fixed from the observed data and $E_{\gamma}$ is the observed VHE photon energy ($E_{\gamma}\simeq 0.1\, E_p$), $\delta=\alpha+\beta$ is the VHE photon spectral index and $\beta$ is the spectral index of the background seed photons in the jet environment. The exponential factor is the survival probability of the VHE gamma-ray due to its interaction with the EBL through the process $\gamma\gamma\rightarrow e^+e^-$ with the energy and redshift dependent optical depth $\tau_{\gamma \gamma}$~\citep{1992ApJ...390L..49S,doi:10.1126/science.1227160,Padovani:2017zpf}. Several EBL models are used to analyze the VHE spectra of extragalactic sources~\cite{Franceschini:2008tp,2010ApJ...712..238F,Dominguez:2010bv,10.1111/j.1365-2966.2012.20841.x,2021MNRAS.507.5144S}. 
The above photohadronic model is used with great success to explain the VHE gamma-ray flaring events from several HBLs. For the HBLs, the background seed photons are always in the low-energy tail region of the SSC spectrum, which follows a perfect power-law as $\Phi_{SSC}\propto \epsilon^{\beta}_{\gamma}\propto E^{-\beta}_{\gamma}$.
It is observed that the photohadronic model fits the VHE spectrum well for 
$E_{\gamma} > 100$ GeV and below this energy the main contribution to the spectrum is due to the leptonic processes.

Mrk 421, Mrk 501 and 1ES 1959+650 are nearby, well known and extensively studied HBLs. Using the photohadronic model we have studied their VHE spectra very well. It is observed that these sources sometimes exhibit transient EHBL-like behavior~\citep{2015A&A...578A..22A,Ahnen:2018mtr,2020A&A...638A..14M} and the standard photohadronic model is unable to explain these behavior. To overcome this difficulty, we proposed that the tail region of the SSC spectrum responsible for the production of $\Delta$-resonance has two different power-laws as~\citep{Sahu:2020tko,Sahu:2020kce,Sahu:2021wue}, 
\begin{equation}
\Phi_{SSC}\propto
 \left\{ 
\begin{array}{cr}
E^{-\beta_1}_{\gamma}
, & \quad 
 100\, GeV\, \lesssim E_{\gamma} \lesssim E^{t}_{\gamma}
\\ E^{-\beta_2}_{\gamma} ,
& \quad   E_{\gamma}\gtrsim E^{t}_{\gamma}
\\
\end{array}
\right. ,
\label{eq:sscflux}
\end{equation}
where $\beta_1\neq\beta_2$ and $E^{t}_{\gamma}$ is the transition energy between two different zones. The value of $E^{t}_{\gamma}$ differs from one spectrum to another and provides information on the nature of the flaring event. Using Eq. (\ref{eq:sscflux}), the photohadronic model is extended to include two different zones as,
\begin{equation} 
F_{\gamma}(E_\gamma)=
e^{-\tau_{\gamma\gamma}(E_\gamma,z)}\times
\begin{cases}
F_1 E_{\gamma , TeV} ^{-\delta_1+3}
,&\hspace{-0.2cm} 100\, \mathrm{GeV}\,\lesssim E_{\gamma} \lesssim E^{t}_{\gamma}\,\, (\text{zone-1}) \\ 
F_2 E_{\gamma , TeV} ^{-\delta_2+3},
& \,\,\,\,\, E_{\gamma}\gtrsim E^{t}_{\gamma}\,\, (\text{zone-2})
 \end{cases},
\label{eq3:flux}
\end{equation}
where the spectral indices are defined as $\delta_i=\alpha+\beta_i$ ($i=1,2$). In zone-1, $2.5 \le \delta_1 \le 3.0$ and $F_1$ is its flux normalization. Similarly, in zone-2, $3.1 \leq \delta_2 \leq 3.5$ and $F_2$ is its flux normalization. Such a two-zone photohadronic model explains very well the VHE spectra of many temporary EHBLs~\citep{2019ApJ...884L..17S,Sahu:2020tko,10.1093/mnras/stac2093}.

Although the EHBL-like behavior of 1ES 2344+514 is exactly the same as other temporary EHBLs, its VHE spectra of 2016 August 11 and 12 have peculiar behavior and can be explained by zone-2 only and zone-1 is absent here, so, no need of $E^{t}_{\gamma}$~\citep{10.1093/mnras/stac2093}. To our knowledge, no other temporary EHBL exhibits this behavior and is probably a new type of event observed for the first time. The VHE spectra of GRB 221009A observed by WCDA and KM2A have the same behavior as the spectra of 1ES 2344+514 observed on 2016 August 11 and 12.


\begin{figure}
\centering
\includegraphics[width=1\linewidth]{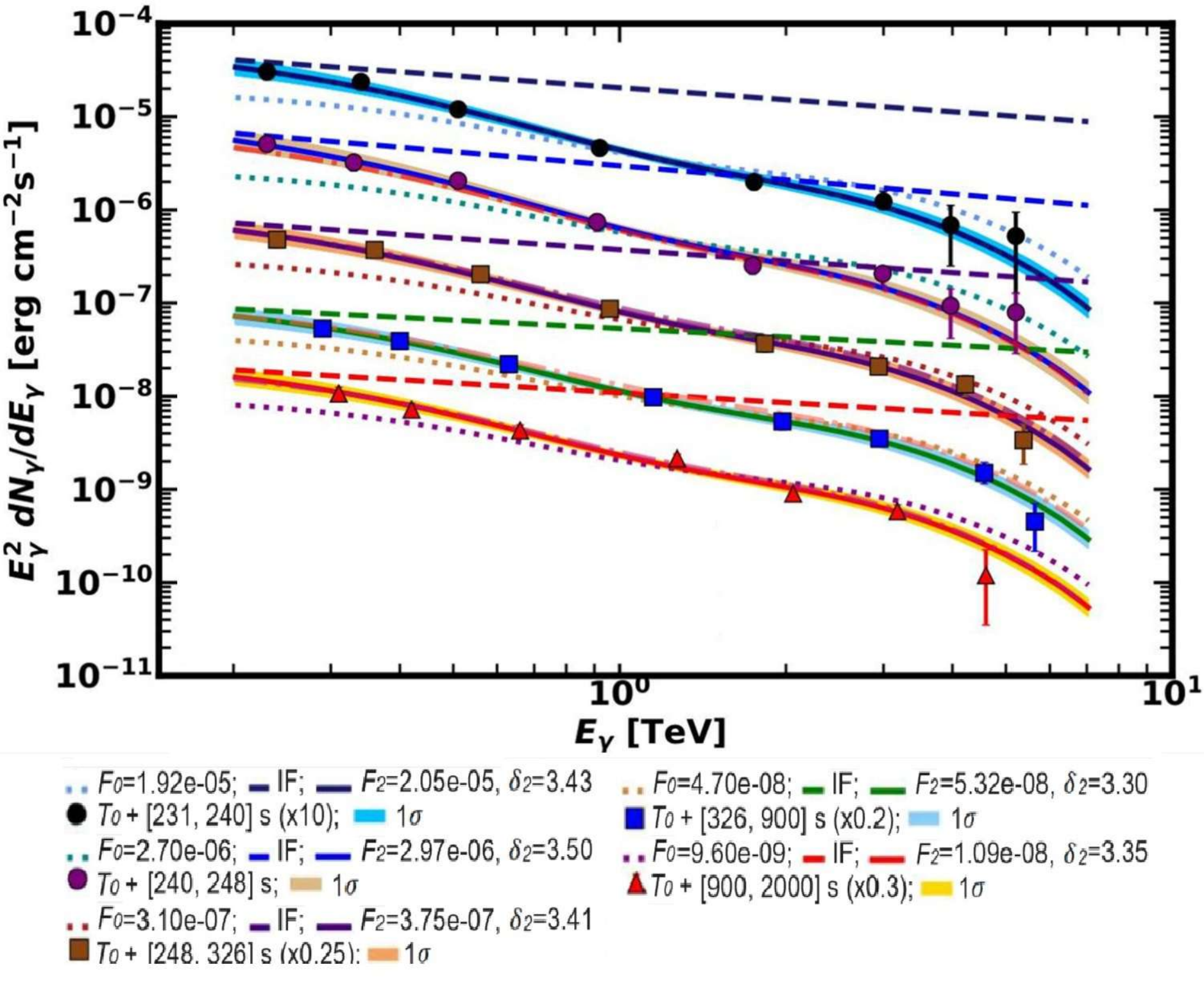}
\caption{The LHAASO-WCDA observation of the VHE gamma-ray flux during the rising phase, the peak phase and three decay phases in different time intervals are fitted with the photohadronic model. In this, and in all the subsequent figures, the values of $F_0$ and $F_2$ are defined in units of $\mathrm{erg\,cm^{-2}\, s^{-1}}$. The shaded region in each spectrum corresponds to the $1\sigma$ confidence interval corresponding to best fit values of $F_2$ and $\delta_2$. We use the same flux scaling as given in~\citep{2023Sci...380.1390L} for comparison. For each spectrum, we have also shown the PLEC fit to the observed spectra (dashed dotted curves) and the values of $A$ (the normalization factors), $E_{cut}$ are taken from Table S2 of~\citep{2023Sci...380.1390L}. The fits with $\delta=3.0$ and $F_0$ are the dotted curves. The intrinsic flux (IF) curves corresponding to each zone-2 fit are the dashed lines. The zone-2 fit (continuous curve) and the PLEC fit   are almost the same.
} 
\label{fig:figure1}
\end{figure}



\begin{figure}
\centering
\includegraphics[width=1\linewidth]{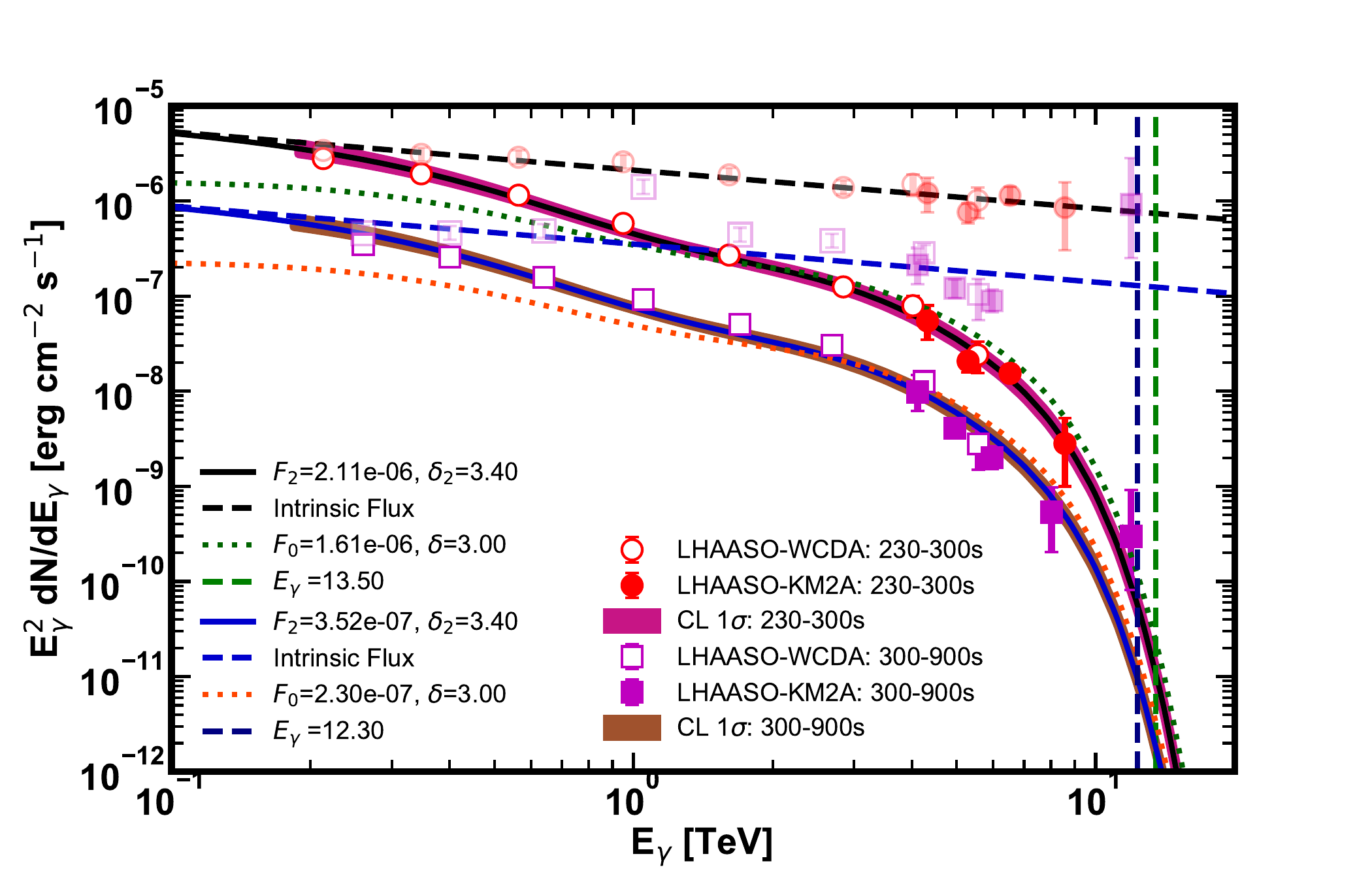}
\caption{The combined SEDs of WCDA and KM2A in the time intervals of $T_0+[230,300]$ s and $T_0+[300,900]$ s are fitted using the photohadronic model with $\delta=3.0$ (dotted curves) and with zone-2 fit (solid curves).  The shaded region in each SED corresponds to the $1\sigma$ confidence interval corresponding to the best fit values of $F_2$ and $\delta_2$. The intrinsic curves are the dashed curves corresponding to each zone-2 fit and their intrinsic data points are also shown. Vertical dashed lines at $E_{\gamma}=13.5$ TeV (green)  and 12.3 TeV (blue) are shown for the upper limits of the photon energy we consider. 
} 
\label{fig:figure2}
\end{figure}



\begin{figure}
\centering
\includegraphics[width=1\linewidth]{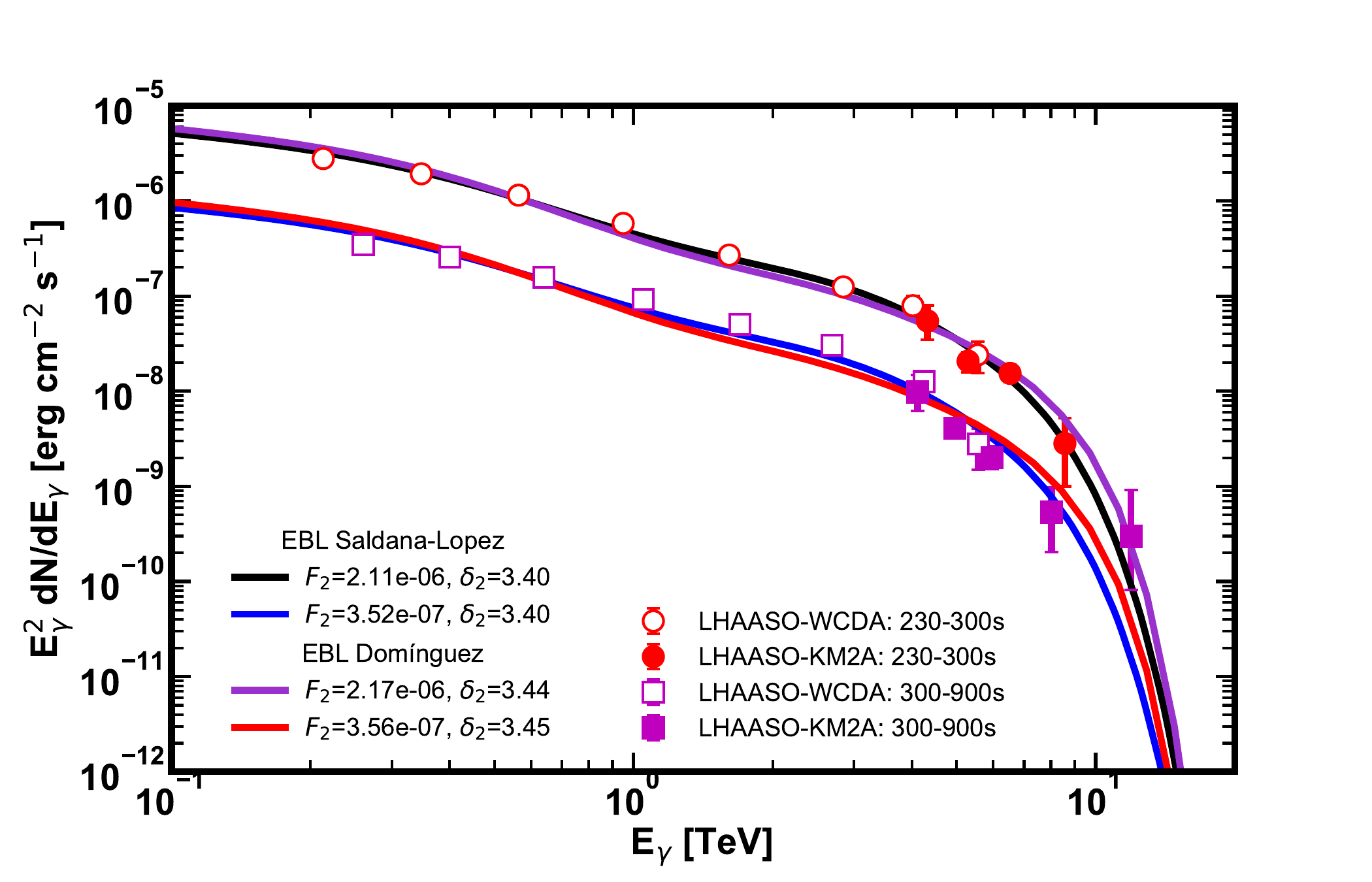}
\caption{The combined SEDs of WCDA and KM2A in the time intervals $T_0+[230,300]$ s and $T_0+[300,900]$ s are fitted with the zone-2 of the photohadronic model accompanied by the EBL Saldana-Lopez and the EBL Domínguez for comparison. 
} 
\label{fig:figure3}
\end{figure}



\begin{figure}
\centering
\includegraphics[width=1\linewidth]{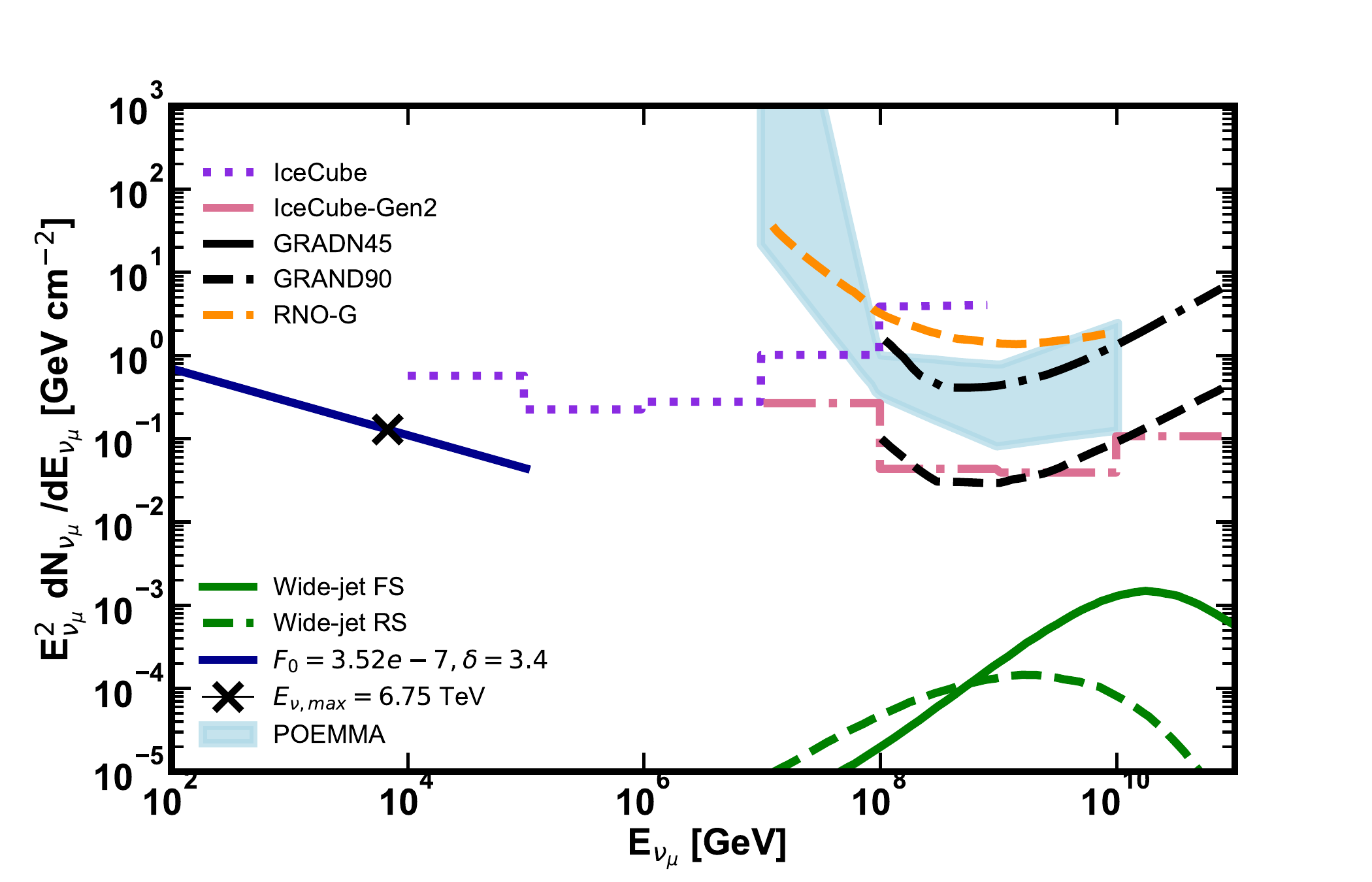}
\caption{The per-flavor time integrated neutrino number flux is estimated in the photohadronic model and compared with the sensitivity of the existing and forthcoming experiments. The cross sign corresponds to the integrated neutrino number flux at maximum neutrino energy, $E_{\nu,\mathrm{max}}=6.75$ TeV, in the photohadronic model. We show the sensitivity curves of the IceCube at a source declination angle $\theta_D=-23^{\circ}$~\citep{IceCube-Gen2:2021rkf} and the forthcoming experiments: IceCube-Gen2 at $\theta_D=0^{\circ}$~\citep{IceCube-Gen2:2021rkf}, GRAND at $\theta_D=45^{\circ}$ and zenith angle $90^{\circ}$~\citep{2020SCPMA..6319501A},POEMMA taking the 90\% unified confidence level and assuming observations during astronomical night~\citep{2020PhRvD.102l3013V}, and RNO-G at $\theta_D=77^{\circ}$~\citep{2021JInst..16P3025A}. We also compare our results with the predictions of reverse shock (RS) and forward shock (FS) proton synchrotron models~\citep{2023arXiv231113671Z}. 
}
\label{fig:figure4}
\end{figure}


\section{\label{sec4}Results and Conclusions}

\subsection{LHAASO-WCDA Spectra}

The LHAASO-WCDA detected more than 64,000 photons from GRB 221009A in the energy range $0.2 \,\mathrm{TeV} \leq \, E_{\gamma} \, \leq 7\, \mathrm{TeV}$ within $T_0+3000$ s. The temporal profile of the flux was smooth with a rapid rise to a peak value and a decay phase (as shown in Fig. 2A of~\cite{2023Sci...380.1390L}). Five time intervals: the rising phase $T_0+[231, 240]$ s, the peak $T_0+[240, 248]$ s, and three decay phases $T_0+[248, 326]$ s, $T_0+[326, 900]$ s, and $T_0+[900, 200]$ s, respectively, were chosen from the observation of LHAASO-WCDA to cover the main afterglow period. In these time intervals, the observed and the intrinsic VHE fluxes are plotted as functions of energy $E_{\gamma}$. It is shown that the TeV afterglow emission can be explained by the SSC process in the external forward shock region. However, these spectra also fitted with PLEC functions. We used the photohadronic model along with the EBL model of~\cite{2021MNRAS.507.5144S} (EBL Saldana-Lopez) to explain these five VHE spectra, and good fits were obtained for $\delta=3.0$, which corresponds to the low emission state. These are shown in Figure~\ref{fig:figure1}. However, flaring of VHE in low emission state is very common in HBLs with $\sim 48$\% probability~\citep{2019ApJ...884L..17S}. GRB221009A is titanic, as it is the brightest and extremely rare GRB ever detected with the highest total isotropic equivalent energy~\citep{Williams:2023sfk}. It seems very unlikely that its VHE spectra, especially the spectra in the initial rising phase and the peak phase can be in the low emission states. Thus, instead of restricting $\delta$ in the range $2.5\le \delta \le 3.0$, we use it as a free parameter in the analysis. 

Using $\delta$ as a free parameter and EBL Saldana-Lopez, excellent fit to the VHE spectra of all five phases of LHAASO-WCDA is obtained for $\delta \gtrsim 3.3$, which are shown in Figure~\ref{fig:figure1}. We have also used the EBL model of~\cite{Dominguez:2010bv} (EBL Domínguez) to fit the spectra and compare with the fit of EBL Saldana-Lopez. It is observed that both the EBL models fit well to the observed spectra. Henceforth, our analysis will be based on the EBL Saldana-Lopez.

According to the classification of BL Lac objects, the present VHE emissions from the afterglow phases of GRB 221009A observed by LHAASO-WCDA belong to the temporary EHBL-like, similar to the one observed in the VHE flaring of 1ES 2344+514 during August 2016, which can be explained with zone-2 only and $\delta$ is replaced by $\delta_2$ with $3.1 \lesssim \delta_2 \lesssim 3.5$ and zone-1 is absent here~\citep{10.1093/mnras/stac2093}. This clearly shows that the VHE gamma-ray spectra from GRB 221009A are not in low emission states but are in temporary EHBL-like states. Like the VHE flaring of 1ES 2344+514, the multi-TeV gamma-ray event from GRB 221009A is also a new type of VHE emission, which we observe for the first time. Probably this is the reason why GRB 221009A is unique and exceptional.

The best fit to the spectra during the rising phase, the peak and the three decay phases are shown in Figure~\ref{fig:figure1} with their respective $\delta_2$ value and all of them satisfy $3.1 \lesssim \delta_2 \lesssim 3.5$. Also, their respective intrinsic fluxes $F_{in}$ are shown in the figure, where $F_{in}\propto E^{-\delta_2+3}_{\gamma}$. We have compared the temporary EHBL-like fit (zone-2 fit)~\citep{10.1093/mnras/stac2093} with the low emission state fit $\delta=3.0$, and the PLEC fit. It can be seen that the PLEC fit is very similar to the zone-2 fit. We have calculated the $1\sigma$ confidence intervals to zone-2 fit to each spectrum which are shown in Figure~\ref{fig:figure1}.

\subsection{Combined SED of LHAASO-WCDA and LHAASO-KM2A}

The LHAASO-KM2A is optimized to observe photons in the energy range of 10 TeV to 10 PeV. However, it can still observe photons below 10 TeV with some uncertainties. Between 230 s to 900 s after $T_0$, LHAASO-KM2A detected 142 events from the direction of GRB 221009A with energies above 3 TeV. Due to low-energy resolution below 10 TeV, LHASSO-KM2A assumed two different spectral functions: LP and PLEC for the reconstruction of energies in each event. Assuming LP function, eight events above 10 TeV were observed and $17.8^{+7.4}_{-5.1}$ TeV was the highest energy~\citep{2022GCN.32677....1H}. On the other hand, using the PLEC function, the highest photon energy was $12.2^{+3.5}_{-2.4}$ TeV. At a redshift of $z=0.151$, the survival probability $e^{-\tau_{\gamma\gamma}(E_\gamma,z)}$ of a photon of $E_{\gamma}=12.2$ TeV is $\sim (8.5\times 10^{-5} - 2.3\times 10^{-4})$, while, for $E_{\gamma}=17.8$ TeV, it is $\sim (7.1\times 10^{-9} - 6.1\times 10^{-9})$, respectively with EBL models of ~\cite{2021MNRAS.507.5144S} and ~\cite{Dominguez:2010bv}. Thus, observation of the photon with $E_{\gamma}\gtrsim 13$ TeV implies a relatively transparent Universe or a signature of new physics~\citep{2022arXiv221007172B,Troitsky:2022xso,Zhu:2022usw,Li:2022wxc,2023ApJ...942L..21F,2022arXiv221005659G,Lin:2022ocj}. 

The evolution of GRB 221009A from the prompt phase to the afterglow phase was observed simultaneously by LHAASO-WCDA in the energy range of 0.2 TeV to 7 TeV and by LHAASO-KM2A above 3 TeV and up to $\sim 13$ TeV. For the spectral analysis, KM2A data were split into time intervals $T_0+[230, 300]$ s and  $T_0+[300, 900]$ s and combined with the WCDA data observed during the same time intervals. In the overlapping region, 3 TeV to 7 TeV, both observations are consistent, indicating a common origin. The combined VHE spectrum in a given time interval should be explained by a single photon spectral index, if the KM2A spectrum has no cutoff. We have seen that the spectra of LHAASO-WCDA are temporary EHBL-like and are explained extremely well with $3.1 \lesssim \delta_2 \lesssim 3.5$. Therefore, by extending these spectra to higher energies, we should also be able to explain the LHAASO-KM2A events.

The combined SED in the time intervals $T_0+[230, 300]$ s and $T_0+[300, 900]$ s is fitted using the photohadronic model by taking $\delta$ and $F_0$ as free parameters. Like the spectra of LHAASO-WCDA, these spectra are also very well fitted with $3.1 \lesssim \delta_2 \lesssim 3.5$, which confirms that the combined spectra are also temporary EHBL-like and only in zone-2~\citep{10.1093/mnras/stac2093}. For $T_0+[230, 300]$ s, the best fit to the spectrum is obtained for $\delta_2=3.40$ and $F_2=2.11\times 10^{-6}\, \mathrm{erg\,cm^{-2}\,s^{-1}}$. Similarly, for the time interval $T_0+[300, 900]$ s, the best fit is obtained with $\delta_2=3.40$ and $F_2=3.52\times 10^{-7}\, \mathrm{erg\,cm^{-2}\,s^{-1}}$. These are shown in Figure~\ref{fig:figure2}, where we have also shown the $1\sigma$ confidence intervals for these curves and compare with the $\delta=3.0$ fits. In Figure~\ref{fig:figure3}, we compare the EBL Saldana-Lopez fit with the EBL Domínguez fit and observe a small difference  above $\sim 6$ TeV.

It is very important to note that in order to fit the observed VHE spectra of WCDA and KM2A, the value of the photon spectral index ($\delta_2$ here) is the only parameter we have to know. The value of $\delta_2$ tells us about the type of VHE event and whether the background photons belong to the synchrotron regime or the SSC regime. The detailed knowledge of $\Gamma$, $\epsilon_{\gamma}$ and any other parameters are not needed. However, to estimate the range of $\Gamma$ we need to know the range of $\epsilon_{\gamma}$. 

For $3.1 \lesssim \delta_2 \lesssim 3.5$, we have $\beta_2 >0$ corresponds to the background seed photons in the low-energy tail region of the SSC SED and the flux is $\Phi_{SSC}\propto \epsilon^{\beta_2}_{\gamma}\propto E^{-\beta_2}_{\gamma}$. The VHE gamma-rays in the energy range $0.2 \,\mathrm{TeV} \leq \, E_{\gamma} \, \leq 13$ TeV are produced from the interaction of protons in the energy range $2 \,\,\mathrm{TeV} \leq \, E_{p} \, \leq 130$ TeV with the SSC background seed photons in the forward shock region of the GRB jet. We can estimate $\Gamma$ from the kinematic condition to produce $\Delta$-resonance. Normally, the low-energy tail of the SSC SED of EHBLs starts $\sim 10$ MeV. By assuming that the seed SSC photons are in the energy range  $10\,\, \mathrm{MeV} \leq \epsilon_{\gamma} \leq\, 250$ MeV, the corresponding $\Gamma$ for the maximum proton energy $E_p=130$ TeV will be in the range $73 \leq \Gamma \leq 367$. Using the minimum variability time scale of $t_v\sim 20$ s~\citep{footnote1} during the afterglow and $\Gamma\sim 367$, the dissipation radius is $R'=2 \Gamma^2 t_v\sim 1.6\times 10^{17}$ cm. The optical depth for $\Delta$ production with the SSC photon background in the dissipation region is $\tau_{p\gamma}=n'_{\gamma}\sigma_{\Delta} R'$, where $n'_{\gamma}$ is the comoving SSC photon density in the forward shock region. To produce the $\Delta$-resonance with a mild efficiency, we impose $\tau_{p\gamma} < 1$ and this gives $n'_{\gamma} < 1.2\times 10^{10}\, \mathrm{cm^{-3}}$. In the same SSC background, the $e\gamma$ interaction will also take place and electrons lose energy. Hence, we should take $\tau_{e\gamma}\, > 1$, which gives $n'_{\gamma} > 9.3\times 10^{6}\, \mathrm{cm^{-3}}$. By taking $n'_{\gamma} \sim  10^{8}\, \mathrm{cm^{-3}}$, we have $\tau_{p\gamma} \sim 8.1\times 10^{-3}$.

For production of $\Delta$-resonance within the jet during the afterglow epoch, the acceleration timescale $t'_{acc}=10\eta E'_p/eB'$ and the $p\gamma$ interaction timescale $t'_{p\gamma}=(\sigma_{p\gamma} K_{p\gamma} n'_f)^{-1}$ in the jet comoving frame should satisfy $t'_{acc}\,<  t'_{p\gamma}$~\citep{2020ApJ...898..103S}. The maximum proton energy $E_p=130$ TeV corresponds to a comoving frame energy $E'_p=0.41$ TeV. By considering a magnetic field $B'\sim 10^{-3}$ G at the dissipation radius $\sim 1.6\times 10^{17}$ cm and by taking $\eta \sim 100$ ~\citep{Cerruti:2014iwa}, which characterizes the magnetic disturbances responsible for the particle acceleration in the jet, we get $t'_{acc}\sim 4.6\times 10^4$ s and $t'_{p\gamma}\sim 3.3\times 10^9$ s. Thus, the condition $t'_{acc}\,<  t'_{p\gamma}$ is satisfied. Recently, by comparing different timescales in the comoving frame of the jet in GRB 221009A, it is shown that ultra-high energy cosmic ray protons can be accelerated $\sim 10^3$ EeV energy~\citep{2024ApJ...963..109H}. 

The $\pi^0$ decays to two VHE photons in the background of seed SSC photons of energy $\epsilon_{\gamma} \gtrsim 10$ MeV. These VHE photons interact with the SSC photons to produce $e^+e^-$ with a cross section of $\sigma_{\gamma\gamma}\, \lesssim 1.3\times 10^{-30}\, \mathrm{cm^{-2}}$. The mean free path of the VHE photons in the SSC photon background is $\lambda_{\gamma\gamma} =(n'_{\gamma}\,\sigma_{\gamma\gamma})^{-1}\,\gtrsim 8\times 10^{21}\,\mathrm{cm}$, $R' \ll \lambda_{\gamma\gamma}$. Thus, the VHE photons will not be attenuated by the SSC photon background.

The Bethe Heitler (BH) pair production process can also contribute to proton cooling. The $e^+e^-$ produced from the BH process will emit synchrotron photons and the maximum energy of these photons is in the low-energy tail region of the SSC spectrum. Thus, the VHE spectrum will not be affected by the BH process~\citep{Petropoulou:2014rla,Petropoulou:2015upa}.

For $3.3 \lesssim \delta_2\lesssim 3.5$, the sum of the integrated flux of LHAASO-WCDA in the five phases in the energy range $0.2 \,\,\mathrm{TeV} \leq \, E_{\gamma} \, \leq 7$ TeV is $F^{int}_{\gamma}\simeq 9.9\times 10^{-6}\, \mathrm{erg\,cm^{-2}\,s^{-1}}$ and the corresponding isotropic luminosity is $L_{iso, TeV}\simeq 6.7\times 10^{50}\,\mathrm{erg\,s^{-1}}$. The $F^{int}_{\gamma}$ for the combined SED of WCDA and KM2A during the time interval $T_0+[230, 900]$ s in the energy range $0.2 \,\,\mathrm{TeV} \leq \, E_{\gamma} \, \leq 13.5$ TeV is $F^{int}_{\gamma}\simeq 3.5\times 10^{-6}\, \mathrm{erg\,cm^{-2}\,s^{-1}}$ and the corresponding isotropic luminosity is $L_{iso, TeV}\simeq 2.3\times 10^{50}\,\mathrm{erg\,s^{-1}}$.

In the photohadronioc model, the decay of charged pions will produce neutrinos. However, nonobservation of neutrinos from GRB 221009A constrain the hadronic models~\citep{2022ApJ...941L..10M,2023ApJ...943L...2L,2023ApJ...944..115A}. The mean lifetime of $\pi^+$ in the jet comoving frame is $\sim 2.6\times 10^{-8}\Gamma\,s$, and before decay, the charged pions may lose substantial amount of energy in the jet environment. Similarly, the $\mu^+$ from the decay of $\pi^+\rightarrow\, \mu^+\nu_{\mu}$ will also lose energy before decay. Thus, the neutrinos produced from pion decay and muon decay have low energies, and may not be detectable by IceCube. Another important point is that the best fit to the VHE spectra are obtained for $\delta_2\simeq 3.4$. So, the intrinsic flux is $F_{in}\propto E^{-0.4}_{\gamma}$. Similarly, the neutrino flux is also $\propto E^{-0.4}_{\nu}$. Thus for higher energy neutrinos the flux will decrease further and may not be detectable by IceCube.

The per-flavor time integrated neutrino number flux $F_{\nu}(E_{\nu})\simeq T\,(F_2/8)\,(E_{\nu}/TeV)^{-0.4}$ can be estimated in the photohadronic model. By taking the total neutrino emission timescale $T \sim 10^4$ seconds and $F_2=3.52\times 10^{-7}\, \mathrm{erg\,cm^{-2}\,s^{-1}}$, we get $F_{\nu}(E_{\nu})\simeq 4.4\, ( E_{\nu}/\mathrm{GeV})^{-0.4}\, \mathrm{GeV\, cm^{-2}}$. The maximum photon energy observed by LHAASO is $E_{\gamma,\mathrm{max}}\sim 13.5$ TeV which corresponds to maximum observed neutrino energy $E_{\nu,\mathrm{max}}(\simeq E_{\gamma,\mathrm{max}}/2)\sim 6.75$ TeV. In the photohadronic model if neutrinos of energy $E_{\nu}\gtrsim 10^4$ GeV are produced during this period, then $F_{\nu}\lesssim 0.11\, \mathrm{GeV cm^{-2}}$, which is smaller than the IceCube sensitivity and also the IceCube-Gen2 sensitivity. The photohadronic model prediction is also compared with other results as shown in Figure \ref{fig:figure4}~\citep{2023arXiv231113671Z}.

In conclusion, we have used the photohadronic model to analyze the VHE gamma-ray spectra observed by LHAASO-WCDA and LHAASO-KM2A from the afterglow of GRB 221009A. The VHE photons are produced from the interaction of high-energy protons with the background SSC photons in the forward shock region of the GRB jet. The VHE SEDs observed by LHAASO-WCDA during different phases and the combined SEDs of both the detectors have temporary EHBL-like behavior, as observed only in the spectra of 11th and 12th of August 2016 from the temporary EHBL 1ES 2344+514, probably a new type of transient EHBL-like source. Also, this is the first time such a temporary EHBL-like emission is observed from a GRB. Additionally, its closeness to the Earth and beaming towards us with an opening angle of~$2/\Gamma\sim 0.3^{\circ}$, probably makes the GRB 221009A a unique and exceptional event.

It is puzzling that the spectra of GRB 221009A do not behave like the spectrum of a temporary EHBL with two-zones which are more in number and have similar properties to that of 1ES 2344+514. In future, observation of  many more GRBs in multi-TeV photons at $z < 1$  can shed more light on the nature of VHE emissions, and we speculate that we should be able to observe some of them with two-zone VHE emissions, similar to the one observed from transient EHBL-like events from several nearby HBLs.

\section*{Acknowledgements}

The work of S.S. is partially supported by DGAPA-UNAM (México) Projects No. IN103522. B. M-C and G. S-C would like to thank CONAHCyT (México) for partial support. Partial support from CSU-Long Beach is gratefully acknowledged. S.S. is thankful to Martin Pohl and Bing Zhang for many fruitful discussions.


\section*{Data Availability}
No new data were generated or analysed in support of this research.


\bibliographystyle{mnras}
\bibliography{grbref2}


\bsp	
\label{lastpage}
\end{document}